# SAT for Pedestrians

Bernd R. Schuh

Dr. Bernd Schuh, D-50968 Köln, Germany; bernd.schuh@netcologne.de

**Abstract.**
The aim of this short note is mainly pedagogical. It summarizes some knowledge about Boolean satisfiability (SAT) and the P=NP? problem in an elementary mathematical language. A convenient scheme to visualize and manipulate CNF formulae is introduced. Also some results like the formulae for the number of unsatisfied clauses and the number of solutions might be unknown.

**Introduction**.

Specifically, I will formulate the problem of finding solutions for SAT in fairly elementary terms. Ordinary school calculus will be sufficient to do most of the calculations. One will see how the exponential character of the assignment space undermines any straightforward attempt of finding an efficient algorithm for the problem. Some new necessary conditions ("checks") for a formula to be satisfiable (or not) will be deduced. Most checks can be calculated in polynomial time("p.t."). Their usefulness has not yet been tested for large formulae.

The search for "efficient algorithms" (i.e. polynomial time procedures), is important for settling the "P=NP?" problem, one of the outstanding problems of 21st century mathematics. For a definition of complexity classes P and NP a vast amount of literature, both in print and online is available. A simple Google search will do more than necessary. The essence is: problems in class P can be solved on a classical Turing computer, and their running times are bounded by a polynomial in the length of the problem (suitably defined). The definition of NP, on the other hand, does not address the <u>solution</u> of an NP problem but is solely interested in the question how long it takes to <u>verify</u> a given solution, i.e. a solution that has been guessed, not determined (also called "certificate"). A problem belongs to NP by definition iff a certificate can be <u>verified</u> or falsified in polynomial time (p.t.).

P=NP states that solutions of problems in NP also can be <u>found</u> in p.t., in principle at least. Whereas P ≠ NP means that some problems in NP need exponential running times for their solution. Those problems are called "hard". (To be precise, a problem is called NP-hard if every problem in NP can be reduced to it in p.t.) Many experts believe that P ≠ NP is true, but a few also believe P=NP, [1]. A proof of the latter could consist in finding an algorithm which solves a generic NP problem in p.t.. But it could also be non-constructive, a pure existence proof. A proof of P ≠ NP would be to find a specific problem which is a member of NP but definitely not of P. Sounds simple, but probably is very difficult, because it amounts to showing that there is no efficient algorithm able to solve the problem. A different route may be to falsify the assumption P=NP by leading it to a contradiction. This appears to be the most interesting approach, because it probably leads into different, possibly new fields of mathematics, where maybe a similar result is already established or easier to prove.

Although the statement P=NP or its negation appear as statements about complexity classes, in essence they are about algorithms as well, and may become obsolete once the concept of Turing machines is abandoned in favor of more advanced concepts. Also the quickly rising power of processors may turn the problem into a purely academic one for all practical purposes very soon. Nevertheless, I will also present some approaches which optimistically aim at finding a p.t. algorithm for NP problems.

Fortunately, for any approach, some classes of NP problems turn out to be generic ("NP-complete"), i.e. finding a p.t. algorithm or disproving that such an algorithm can exist for one of them will settle the question P=NP? once and for all. A problem is called NP-complete iff any element of NP can be reduced to it in p.t.. In the following I will focus on one of these generic problems, namely SAT.

**SAT.**

One of the NP-complete problems is Boolean satisfiability, shortly SAT. The satisfiability problem is to find out whether a given Boolean formula F can be evaluated to "true" under suitable assignments of "true" and "false" to their basic variables. In practical problems (if formulated as SAT problems ) it is more important to find the specific assignments which evaluate the formula to true. They "solve" the problem under consideration. In the following the term "solution" will be used for a fulfilling assignment to a particular formula ("instance").

Note that there can be a difference in finding solutions and in proving that one exists. The difference stems from the dimension of the solution space. For a formula with n basic variables there are $2^n$ different assignments. The dimension of the solution space (i.e. the cardinality of the set of all solutions) may well be of that order. Therefore simply enumerating all solutions is impossible in polynomial time, save from finding them. In other words, the complexity of the Boolean satisfiability problem does not reside in the complexity of the solution space (whatever the definition of that may be), as is known since 1986 by the Valiant–Vazirani theorem [2]. The theorem implies that the Boolean satisfiability problem remains a difficult problem with high complexity (a "hard" problem), even if only formulae are taken into account which have at most one solution. Therefore claims that the "hardness" of a problem is associated with the complexity of the solution space, [3], are to be taken with skepticism.

To establish P=NP it suffices to prove that a p.t. algorithm exists which decides whether a given formula has a solution at all. For that purpose one should be aware that each instance is the representative of the class of logically equivalent formulas to which it belongs. This might appear trivial but should be kept in mind. Furthermore there are simple symmetry operations, like e.g. flipping variables from their positive to their negative form, which change the appearance of a formula but conserve the number of solutions and thus its satisfiability. Such transformations do not change the complexity of the problem or the structure of the solution space (if it is considered as a metric space with Hamming distance).

**CNF- formulae and scheme-representation.**

In the following the focus will be on propositional formulae in conjunctive normal form, CNF. I will not use the set theoretical description, see e.g. [4], which stresses the interchangeability of clauses and of variables as well. Instead I write such formulae in the common form

$$(1) \quad F = C_1 \wedge C_2 \wedge ... \wedge C_m; \quad C_i = l_{i1} \vee l_{i2} \vee ... \vee l_{in}$$

with m clauses $C_i$ and n basic variables $a_j$. The $l_{ij}$ are used as fil-ins for either

a variable $a_j$, or

its negation $\bar{a}_j$, or

falsity $\bot$, representing any statement which is false under all truth assignments, or absolute truth T, representing any tautology, being true under all assignments by definition.

The first index – i - refers to the clause number. Note that tautologies cannot be written in CNF form (1) unless T is allowed as a fil-in.

A new short hand notation will be convenient, in particular to visualize a CNF formula. Write F as a (m,n)-matrix-scheme where rows represent clauses, columns variables. At position (i,j) we write

$$\text{"matrix} - \text{scheme"} \quad \begin{array}{ll} x & \text{iff clause i contains } a_j \\ \bar{x} & \text{iff clause i contains } \bar{a}_j \\ 0 & \text{iff clause i contains } \varnothing \end{array}$$

To give an example, the scheme

$$F_4 = \begin{matrix} 0 & \bar{x} & x & \bar{x} \\ x & \bar{x} & \bar{x} & 0 \\ \bar{x} & 0 & \bar{x} & 0 \\ 0 & x & 0 & 0 \end{matrix}$$

represents the formula

$$F_4 = (\bar{a}_2 \vee a_3 \vee \bar{a}_4) \wedge (a_1 \vee \bar{a}_2 \vee \bar{a}_3) \wedge (\bar{a}_1 \vee \bar{a}_3) \wedge (a_2) \ .$$

**Straightforward determinations of satisfiability**.

There are trivially solvable CNF formulae which can be identified at a glance on the representing matrix scheme. We state a first example as a lemma:

*Lemma Ia*
If each row contains at least one x, F is satisfiable.

This is apparent since setting all variables to "true" will fulfill F. Similarly,

*Lemma Ib*
If each row contains at least one $\bar{x}$ , F is satisfiable.

In this case all variables set to "false" will be a fulfilling assignment.

Obviously, $F_4$ is not of that kind, because of rows 3 and 4. On the other hand, $F_4$ shows that the converse is not true: A scheme that contains rows with only one variety of the variables and one or more rows with the other variety, will not necessarily be unsatisfiable. I will call such rows "only-x-rows" and "only-$\bar{x}$-rows". $F_4$ for instance is such a formula, row 3 being an only-$\bar{x}$ –row, and row 4 an only-x-row. But $F_4$ has two solutions. Note that testing a formula for "only-$\bar{x}$-rows" and "only-x-rows" obviously is a p.t. test. The test is not sufficient for establishing non-satisfiability, only necessary.

This becomes clear by considering the equivalence class of a CNF formula which is generated by performing all possible variable "flips" $a_i \leftrightarrow \bar{a}_i$, or in terms of the scheme representation by interchanging $x \leftrightarrow \bar{x}$ in one or more columns. I will call this class FLIP(F) in the following.

For a CNF formula F with n variables and m clauses FLIP(F) contains $2^n$ formulae with the same number of variables and clauses. Moreover, all elements of FLIP(F) have the same number of solutions, that is the number of solutions is an invariant under the flipping operation, [5]. In particular all members of FLIP(F)  are either satisfiable or non-satisfiable. That is so because any solution of F can be transformed into a solution of a different member of FLIP(F), F', simply by switching the values from "true" to "false" and vice versa for those variables which have been flipped in going from F to F'. One may also establish a one-to-one correspondence between the elements of FLIP and the possible assignments.

Consider $F_4$ as an example. Its solutions correspond to flipping columns 1,3 and 4 simultaneously, or 3 and 4 simultaneously. Both operations transform $F_4$ into new forms which fulfill the assumption of lemma I, thus they are trivially solved by setting all variables to true. Thus going back from the transformed formulae to $F_4$ generates the two solutions (false, true, false, false) and (true, true, false, false).

Consider, however, the enlarged formula $F_5$, given by

$$F_5 = \begin{array}{cccc} 0 & \bar{x} & x & \bar{x} \\ x & \bar{x} & \bar{x} & 0 \\ \bar{x} & 0 & \bar{x} & 0 \\ 0 & x & 0 & 0 \\ 0 & 0 & 0 & x \end{array}$$

where an additional clause $a_4$ has been added. Now none of the aforementioned operations transforms $F_5$ into a form which is solvable by (true,true,true,true). In fact, there is no flipping operation at all that does the trick. Thus no member of FLIP($F_5$) is solved by that assignment, therefore $F_5$ is unsatisfiable. To get this result all $2^n$ (=16 in the example) possible operations (or members of FLIP) have to be checked, in principle.

Flipping conserves the form (i.e. m and n do not change) and the number of solutions. It does not conserve the solutions itself. There are other operations which lead to logically equivalent formulae in this stronger sense, i.e. formulae with identical solutions. I will use the symbol $\triangleq$ for this equivalence. Trivial manipulations of this sort are permutations of rows and columns. In the first case the equivalence is a consequence of the commutativity of the $\wedge$ operation. In the second case, the same formula is restored by a renumbering of variables.

Other operations which keep a formula in the same equivalence class do not necessarily conserve the number of clauses, m. One is the "blow up" or "shrink" operation, given by

$$\text{"blow up"/"shrink"} \quad \begin{array}{ccc} R & x & S \\ R & \bar{x} & S \end{array} \triangleq R \; 0 \; S$$

where R and S represent any left and right parts of a row. It is a consequence of the distributive laws for logical operations, (or $(X \vee A) \wedge (\bar{X} \vee A) \triangleq A$ for any formulae X and A). It works in both directions. Blow up is particularly interesting because it allows to eliminate all zeros in a scheme and "complete" each clause to a "prime". ("primes" are rows without zeros, i.e. clauses which contain all variables, either negative or positive. For a more precise definition see [5].) It is not difficult to see the following lemma:

*Lemma IIa*

F is unsatisfiable iff a fully blown up CNF formula F contains all possible primes (possibly some of them several times).

One can also show the more general statement,[5]:

*Lemma IIb*

The number of solutions of a CNF formula F equals the number of missing primes in the fully blown up version of the formula.

Since there are $2^n$ different primes for a formula with n variables, the implementation of this fact into a search algorithm leads to an exponentially growing search time.

Some simplifications of schemes should be mentioned which shorten the formula without branching. One is "dropping clauses" in the scheme representing the formula:

"dropping clauses"  $\begin{matrix} R & 0 & . & . & . & 0 \\ R & S & . & . & . & . \end{matrix}$  $\triangleq$  $R \quad 0 \quad . \quad . \quad . \quad 0$

Here R and S may be any fillings with x, $\overline{x}$ and zeros, and dots in the upper clause on the lhs and in the clause on the rhs indicate zeros, whereas the dots in the lower clause on the left simply stand for an extension of S. The "Dropping clauses" rule simply expresses the fact $R \wedge (R \vee S) = R$ for any R and S. The rule can also be deduced from blowing up the upper clause on the left up to the point where the lower clause is generated. Which thus becomes superfluous. Note that it is not important where R is located in the formula because by interchanging columns it can always be moved anywhere without changing the solubility properties of the formula.

Another simplification is

"remove only-x-columns":  If a column contains x and zeros only ($\overline{x}$ and zeros only) drop all rows which had an x ($\overline{x}$) in this column and drop this column.

The remaining formula has one variable less. It is satisfiable iff the original was. The operation corresponds to setting the variable of the column to "true" ("false") and searching a solution for the rest of the formula with one variable removed. Thus the "remove only-x-columns" rule is a special case of an assigning operation: If a truth value is assigned to a variable $a_j$ (true/false) then drop column j, and drop all clauses with x/$\overline{x}$ in column j.

Consider $F_5$ as an example. Assigning "true" to variable $a_1$ in $F_5$ leaves the scheme

$\overline{x} \quad x \quad \overline{x}$
$0 \quad \overline{x} \quad 0$
$x \quad 0 \quad 0$
$0 \quad 0 \quad x$

to be solved. Again assigning "true" to variable $a_2$ (which now is represented by the first column), then to $a_3$ leads to

$\begin{matrix} x & \overline{x} \\ \overline{x} & 0 \\ 0 & x \end{matrix} \rightarrow \begin{matrix} 0 \\ x \end{matrix}$

Since now one row is $\emptyset$, the chosen assignments gives no solution. To establish unsatisfiability for the original formula, however, one would have to go back to $F_5$ and also try out "false" for $a_1$, $a_2$, $a_3$ and so forth. Thus in general, the assigning operation is a branching point in any algorithm that uses it, and therefore a potential source of exponential running times.

One last operation to simplify schemes is the treatment of unit clauses (also called "facts"), i.e. rows which, besides from zeros, contain just one x or $\bar{x}$. We may call the operation "accept facts". "Facts" force the corresponding variable to be assigned "true" or "false", respectively. Thus the rule for "accepting facts" is the same as for the assigning operation. But they simplify the formula without branching. E.g. assigning "true" to variables $a_4$, then $a_2$ in $F_5$ leads to

$$
\begin{array}{cccc}
0 & \bar{x} & x & \bar{x} \\
x & \bar{x} & \bar{x} & 0 \\
\bar{x} & 0 & \bar{x} & 0 \\
0 & x & 0 & 0 \\
0 & 0 & 0 & x
\end{array}
\rightarrow
\begin{array}{ccc}
0 & \bar{x} & x \\
x & \bar{x} & \bar{x} \\
\bar{x} & 0 & \bar{x} \\
0 & x & 0
\end{array}
\rightarrow
\begin{array}{cc}
0 & x \\
x & \bar{x} \\
\bar{x} & \bar{x}
\end{array}
$$

and after applying the shrink rule to the two bottom rows one arrives at the contradiction $F \triangleq \begin{array}{c} 0\,x \\ 0\,\bar{x} \end{array}$.

We can state generally: After applying a rule that produces a row of zeros one can stop further manipulations because the remaining formula is unsatisfiable. The same holds if a

"contradiction"  $\begin{array}{cc} x & 0... \\ \bar{x} & 0... \end{array}$

with any number of zeros (to the right or the left of $x/\bar{x}$) is reached. Satisfiability is reached when the result of a manipulation is a

"confirmation"    x  0...    or    $\bar{x}$  0...

**Splitting and Resolution**

A common method to deduce non-satisfiability of a formula is "resolution". Resolution makes use of the fact that the implication $(X \vee Y) \wedge (\bar{X} \vee Z) \Rightarrow Y \vee Z$ is a tautology

(2a)    $(X \vee Y) \wedge (\bar{X} \vee Z) \Rightarrow Y \vee Z$    = T

(for any formulae X,Y,Z). In the resolution method one replaces two clauses of the form on the left of the arrow (if X is one of the basic variables) by the term on the right:

(2b)    "resolution"    $(a_k \vee Y) \wedge (\bar{a}_k \vee Z) \wedge R$    replace by    $(Y \vee Z) \wedge R$

Should $(Y \vee Z) \wedge R$ turn out to be false then also the replaced formula must be false (otherwise the implication cannot be true, which is necessary since it is a tautology). So if repeating this process for all n variables leads to a contradiction, then the original formula must be unsatisfiable. Since the process has at most n steps and formulae get shorter and shorter on the way, it is a p.t. method for determining unsatisfiability. It works well, if the only appearance of the variables $a_k$ is of the form given in (2b) in each step, i.e. all other clauses summarized in R do not depend on $a_k$ or $\overline{a}_k$ anymore. Resolution is inconclusive, however, if that is not the case and the endresult is satisfiable. In this case the original formula may be either satisfiable or not. The effectivity of this method may depend on a clever choice of the order of "resolved" variables.

Take $F_5$ as an example. If first $a_1$ is resolved then $a_4$, then $a_3$ one arrives at a contradiction and one may conclude that $F_5$ is unsatisfiable. Starting with $a_2$ and clauses 2 and 4, then resolving $a_3$ and clauses 1 and 3 does not lead to a contradiction.

Resolution can be considered as a special case of splitting variables. This means: A formula F is satisfiable iff either F with one of the variables set to T (true) is satisfiable or F with the same variable set to $\bot$ (false) is satisfiable, or both are. In schemes

$$
\text{"splitting"} \quad
\begin{array}{cc}
x & Y_1 \\
. & . \\
. & . \\
x & Y_r \\
\overline{x} & Z_1 \\
. & . \\
. & . \\
\overline{x} & Z_s \\
0 & R_1 \\
. & . \\
. & . \\
0 & R_t
\end{array}
\quad \triangleq \quad
\begin{array}{c}
Z_1 \\
. \\
. \\
Z_s \\
R_1 \\
. \\
. \\
R_t
\end{array}
\quad \vee \quad
\begin{array}{c}
Y_1 \\
. \\
. \\
Y_r \\
R_1 \\
. \\
. \\
R_t
\end{array}
$$

Note that r and s can be different. If r=s =1, then splitting reduces to resolution, but with equivalence instead of a one way replacement. In this case the resolution step is exact and conclusive.

Splitting eliminates a variable and can be repeated with a different variable once the right hand side has been brought into CNF form. That is possible by building new rows from all possible combinations $Y_i \vee Z_j$. Not all such rows may survive. For instance they can be neglected if $Y_i$ and $Z_j$ are "orthogonal". That means

> *Definition*
> rows Y and Z are called "orthogonal" iff $Y \vee Z \triangleq T$.

This is the case whenever one row has x, the other $\overline{x}$ in the same column. In e.g. $F_5$ the first clause (row) is orthogonal to all other rows, the second to all except the last one, the third and the fourth row only to the first two, the last row only to the first.

For a set of clauses I will call a subset "non-orthogonal" iff no two clauses of the subset are orthogonal. For determining satisfiability the notion of non-orthogonal subsets can be decisive, as we will see later.

Splitting consists of n steps only and leads to a conclusive result. But the number of clauses increases in each step (from r+s+t to rs+t). Some or many of the $Y_i \vee Z_j$ might be orthogonal and thus negligible in the next step.

Also other simplifications like dropping clauses or removing only-x-columns might simplify the formula resulting from splitting. But there is no guaranty that the number of clauses remains of the same order. In principle it can grow exponentially. Therefore splitting is not a p.t. process.

In essence splitting is identical to building the "metavariable"

$$(3) \quad M_F = \bigvee_{F_i \in FLIP(F)} F_i$$

step by step from flipping one variable after the other, [6]. The result is either T or $\perp$ corresponding to satisfiability or non-satisfiability of F.

The construction (3) is illustrated with $F_5$

$$\begin{matrix} 0 & \bar{x} & x & \bar{x} \\ x & \bar{x} & \bar{x} & 0 \\ \bar{x} & 0 & \bar{x} & 0 \\ 0 & x & 0 & 0 \\ 0 & 0 & 0 & x \end{matrix} \triangleq \begin{matrix} \bar{x} & \bar{x} & 0 \\ \bar{x} & x & \bar{x} \\ x & 0 & 0 \\ 0 & 0 & x \end{matrix} \triangleq \begin{matrix} \bar{x} & 0 \\ x & \bar{x} \\ 0 & x \end{matrix} \triangleq \begin{matrix} \bar{x} \\ x \end{matrix} \triangleq \perp$$

**Special classes of SAT: k-SAT, 2-SAT, HORNSAT, READ-3**

Investigations on Boolean satisfiability are often reduced to examining k-SAT which is the class of CNF formulae with at most k variables (or their negatives) in every clause. k-SAT is NP-complete für k≥ 3. Many investigations focus on k-SAT with k=3. In addition there is a p.t. algorithm to reduce any k-SAT formula to READ-3 form. By definition, in READ-3 formulae each variable (including its complement) appears at most three times in the formula, in other words: in the scheme representing F there are at most 3 non-zero entries (at least m-3 zeros) in each column.

$F_4$ and $F_5$ are examples of a 3-SAT READ-3 formula. To see that just count the x and $\bar{x}$ in each row and each column.

Not all CNF formulae are hard, of course. There are rather trivial subclasses which are solvable in p.t.. Two of them have been introduced in lemmas Ia and Ib already. Another example is 2-SAT. A sketch of a p.t. algorithm to test satisfiability of 2-SAT instances in terms of the schemes is as follows:

1  Remove all only-x-columns and only-$\bar{x}$-columns. If the result is a contradiction/confirmation stop and edit "unsatisfiable"/"satisfiable". Otherwise proceed with 2.

- 2     Accept all facts, successively. Again, if the result is a contradiction or a confirmation stop and edit "unsatisfiable" or "satisfiable". Otherwise the remaining formula contains no more facts and no more only-x/$\bar{x}$ -columns. Store that formula and choose one of the remaining variables. Proceed with 3.
- 3     If you reach this step for the first time assign the chosen variable to "true" and go to 4. Otherwise assign "false" to the chosen variable and go to 5.
- 4     At least one new fact turns up (otherwise there had been an only-x-column). Accept it. Again a new fact turns up. And so on. If the process leads to a confirmation stop and edit "satisfiable". If the process leads to a contradiction go to 3.
- 5     Accept all facts successively. If the process leads to a confirmation stop and edit "satisfiable". If it leads to a contradiction stop and edit "unsatisfiable".

Since all operations reduce the number of variables by at least one, the process comes to a halt. Since the procedure contains only one branching point the algorithm obviously is p.t.

Another example of a subclass of 3-SAT which is in P is called HORNSAT. HORNSAT formulae consist by definition solely of clauses which contain at most one positive literal, i. e. a variable in its positive form, (i.e. x). Thus any scheme representing a HORN formula H is built from only two types of rows: (i) those which all contain at least one $\bar{x}$, and (ii) facts with positive literals. Therefore a procedure to solve H in p.t. looks as follows

- 1     If there are no facts, return H as satisfiable (since it is trivially solved by setting all variables to "false").
- 2     If there are facts in H, accept them all, until a confirmation, contradiction or a formula of type (i) is reached.

Since the HORN property is conserved under accepting facts, and since accepting facts is a p.t. procedure, HORNSAT is solvable in polynomial time.

**Satisfiability as a nonlinear problem.**

The matrix-scheme representation of CNF formulae is helpful in formulating satisfiability in terms of real numbers. First we derive a system of nonlinear equations from which the solutions of a formula F or its unsatisfiability can be derived. Instead of x, $\bar{x}$ in the matrix-scheme (or $l_{ij}$ in equ. (1)) we define

$$f_{ij} = \begin{cases} 1 & \text{iff clause i contains } a_j \\ -1 & \text{iff clause i contains } \bar{a}_j \\ 0 & \text{iff clause i contains } \emptyset \end{cases}$$

Let $T_\mu$ be an arbitrary truth assignment $T_\mu:\{a_1,..., a_n\} \rightarrow \{0,1\}$. By convention we associate 0 with "false", 1 with "true". In terms of the elementary T(a) we have the arithmetic rules:

$$T_\mu(a \vee b) = T_\mu(a) + T_\mu(b) - T_\mu(a)T_\mu(b)$$
$$T_\mu(a \wedge b) = T_\mu(a)T_\mu(b)$$
$$T_\mu(\bar{a}) = 1 - T_\mu(a)$$

Thus from equ. (1) :

$$(4) \quad T_\mu(F) = \prod_{i=1}^{m} T_\mu(C_i) = \prod_{i=1}^{m}[1 - T_\mu(\bar{C_i})] = \prod_{i=1}^{m}[1 - \prod_{j=1}^{n} T_\mu(\bar{l_{ij}})]$$

It is straightforward to relate $T(\bar{l_{ij}})$ to the representation $(f_{ij})$:

$$T_\mu(\bar{l_{ij}}) = (1 - \frac{1}{2}f_{ij}^2)(1 - f_{ij}x_j) \quad \text{where} \quad x_j := 2T_\mu(a_j) - 1 \in \{-1, 1\}.$$

Thus any of the $2^n$ possible assignments for a formula with n variables is represented by a vector $(x_1, x_2, ..., x_n) = x \in \{-1, 1\}^n$. From now on we label different truth assignments by x (not to be confused with the symbol x in scheme representations!) and rewrite (4):

$$(5) \quad T_x(F) = \prod_{i=1}^{m}[1 - c_i g_i(x)] \quad \text{with} \quad g_i(x) = \prod_{j=1}^{n}(1 - f_{ij}x_j) \quad \text{and} \quad c_i = \prod_{j=1}^{n}(1 - \frac{1}{2}f_{ij}^2)$$

Each $c_i$ is a nonzero constant independent of x. Because of the definition of the $f_{ij}$ it is a measure of the number of positive and negative literals in a clause. For a clause with k nonzero literals, e.g.: $c_i = 2^{-k}$. From (5) we can express satisfiability of a formula F by a set of nonlinear equations in the basic truth assignments $T(a_j)$ or the $x_j$, since :

(6)     $T_x(F) = 1$     iff     $g_i(x) = 0$ for all i=1,..., m.

Any x that satisfies $T_x(F)=1$ is a solution. The set of nonlinear equations in (6) yields all solutions, in principle. For m ≫ n the system is over-constrained and one expects F to be unsatisfiable. And vice versa for m ≪ n.

The determination of satisfiability by means of nonlinear equations has been utilized recently to construct new SAT solvers with considerable success, [7].

**Counting solutions**.

There are several ways to determine satisfiability from (6). One is to count solutions:

$$N(F) = \sum_x T_x(F)$$

is the number of solutions of F, if the sum is performed over all possible assignments. Perform the multiplication in $T_x(F)$ (equ. (5)) to get a sum over clusters of clauses

$$(7a) \quad N(F) = \sum_x (1 - \sum_i c_i g_i + \sum_{<ij>} c_i c_j g_i g_j - \sum_{<ijk>} c_i c_j c_k g_i g_j g_k + ...)$$

If any two clauses in one of the higher clusters are orthogonal the contribution of that cluster vanishes because it contains a factor with $f_{is}=-f_{js}$, and thus $g_i g_j = ...(1-f_{is}x_s)(1+f_{is}x_s)... = ...(1-f_{is}^2 x_s^2)... = 0$. Thus only non-orthogonal subsets of clauses contribute to the sum. The x-summation can be performed with the result

$$(7b) \quad N(F) = \sum_{p \in P_m} (-1)^{|p|} [\prod_{s=1}^n \delta(\sum_{j \in p} |f_{js}|, \sum_{j \in p} f_{js})] 2^{\sum_{s=1}^n \delta(0, \sum_{j \in p} |f_{js}|)}$$

(Here $P_m$ is the power set of {1, 2, ..., m}. $\delta(x,y)$ is the Kronecker symbol, giving 1 for x=y and 0 otherwise. Also $\prod_{j \in \emptyset}(\text{anything}) = 1$ and $|\emptyset| = 0$ is understood)

The formula looks a bit complicated, but is easy to deduce and to interpret: Find all non-orthogonal subsets of the set of all clauses of F ( The clumsy $[\prod_{s=1}^n \delta(\sum_{j \in p} |f_{js}|, \sum_{j \in p} f_{js})]$ expresses that condition).

Then in each such subset count the number of columns which consist of zeros only, $\sum_{s=1}^n \delta(0, \sum_{j \in p} |f_{js}|) =: \nu$, and add, if the cardinality of the subset is even, or subtract, if it is odd, $2^\nu$ successively. The first term in the sum, corresponding to p=$\emptyset$ always gives $2^n$. Instead of the number of zeros one can count the number of different variables (x in the scheme) or their negatives ($\bar{x}$ in the scheme) in the subset considered. Then instead of $2^\nu$ one adds or subtracts $2^{n-k_c}$ where $k_c$ is the number of different literals in cluster c. The formula simplifies to

$$(7c) \quad N(F) = 2^n (1 + \sum_c (-1)^{|c|} 2^{-k_c})$$

with $\sum_c$ running over all non-orthogonal clusters (subsets) of clauses.

To illustrate the method, consider $F_5$. The series starts with $2^4$=16. Next the single clauses contribute to the sum with a minus sign: $-2^1-2^1-2^2-2^3-2^3$=-24. Now find all non-orthogonal pairs of clauses. There are four

$$\begin{matrix} x & \bar{x} & \bar{x} & 0 \\ 0 & 0 & 0 & x \end{matrix} \quad , \quad \begin{matrix} \bar{x} & 0 & \bar{x} & 0 \\ 0 & x & 0 & 0 \end{matrix} \quad , \quad \begin{matrix} \bar{x} & 0 & \bar{x} & 0 \\ 0 & 0 & 0 & x \end{matrix} \quad , \quad \begin{matrix} 0 & x & 0 & 0 \\ 0 & 0 & 0 & x \end{matrix}$$

They contribute $2^0+2^1+2^1+2^2$=9. From the last three pairs it is apparent that there is also a triple of mutually non-orthogonal clauses

$$\begin{matrix} \bar{x} & 0 & \bar{x} & 0 \\ 0 & x & 0 & 0 \\ 0 & 0 & 0 & x \end{matrix}$$

yielding $2^0=1$, with a minus sign, since the number of clauses in this subset is odd. Altogether one gets $N(F_5)=16-24+9-1=0$. That is $F_5$ has no solution, it is non-satisfiable.

The first two terms of the sum for N(F) may be written $N(F) = 2^n - \sum_{i=1}^{m} 2^{n-k_i} + ...$ where $k_i$ is the number of literals (variable or negative variable) in clause i. If the rest is ordered with respect to |p| it is clear that in two successive subsets the larger one contributes less in absolute value, because the smaller one contains all clauses that appear in the larger one and with equal or higher powers of 2. Also, the smaller one may contain clauses that do not appear in the larger one. So the sum of all terms beyond the first two terms is larger or equal to zero. So one has

$$(8) \quad N(F) \geq 2^n - \sum_{i=1}^{m} 2^{n-k_i}$$

from which the well known result

*lemma III:*

If $\sum_{i=1}^{m} 2^{-k_i} < 1$ then F is satisfiable

follows. In particular for 3-SAT with all $k_i=3$ one needs at least 8 clauses to build a non-satisfiable formula. There is a simpler proof for the inequality (8), given in the next section.

**Satisfiability and the number of unsatisfied clauses.**

A different way to determine satisfiability from (6) reformulates the problem as a minimization problem. We define, for fixed F, a function u: $\{-1,1\}^n \rightarrow \mathbb{N}\cup\{0\}$ by

$$(9) \quad u(x) = \sum_{i=1}^{m} c_i g_i(x) .$$

Note that u(x) is just the second term in the expansion (7a). u(x) has a meaning: If clause i is unsatisfied by x, then according to (5) $c_i g_i(x)=1$ and vive cersa. Therefore

*Lemma IV*:

u(x) counts the number of clauses not satisfied by x.

Since by definition all $g_i(x) \geq 0$ and $c_i>0$, one has: u(x)= 0 iff all $g_i(x)=0$, and u(x) > 0 iff at least one $g_i(x) \neq 0$. Hence the following holds

*Lemma Va*:

Let F be defined by the $f_{ij}$ and u(x) according to (5), then:

  F is satisfiable    iff    there exists $x \in \{-1,1\}^n$ with u(x) = 0

and its negation

*Lemma Vb*:

F is non-satisfiable    iff    u(x) > 0 for all $x \in \{-1,1\}^n$.

Since u vanishes for all solutions, and for the $2^n$-N(F) non-solutions $u(x) \geq 1$ holds, one has

$$\sum_x u(x) \geq 2^n - N(F) \text{ or}$$

$$N(F) \geq 2^n - \sum_x u(x) = 2^n - \sum_i c_i \sum_x g_i(x) = 2^n(1 - \sum_i c_i)$$

which is exactly condition (8).

Because of lemma IV the definition of u(x) appears to be the natural choice for representing satisfiability as a minimization problem. However, lemma V will hold for <u>any positive coefficients</u> $\alpha_i$ instead of $c_i$ in (9). To take into account the individuality of each clause one might choose different weight factors, for example

$$\alpha_i = c_i 2^{-\left|\sum_{s=1}^n f_{is}\right|}.$$

Also completely arbitrary choices are possible and preserve lemma V, as long as all $\alpha_i > 0$.

**Checks for 3-SAT based on u(x).**

For 3-SAT there are no more than three nonvanishing $f_{ij}$ in each clause i. Thus by expanding the products in (5) and performing the summation over clauses term by term we get from (5) and (9) for 3-SAT formulas

(10a)   $$u(x) = C - \sum_{j=1}^n \lambda_i x_i + \sum_{<ij>} \mu_{ij} x_i x_j - \sum_{<ijk>} \nu_{ijk} x_i x_j x_k$$

where $C, \lambda, \mu, \nu$ are defined by

(10b)   $$C = \sum_{l=1}^m c_l \quad \lambda_i = \sum_{l=1}^m c_l f_{li} \quad \mu_{ij} = \sum_{l=1}^m c_l f_{li} f_{lj} \quad \nu_{ijk} = \sum_{l=1}^m c_l f_{li} f_{lj} f_{lk}$$

and the brackets <> indicate that the sums in (10a) count each pair or triple just once. Note that $\mu, \nu$ are totally symmetric in all indices, and vanish, if any two indices are equal. Therefore $\mu$ is a (n,n)- matrix with at most n(n-1) nonvanishing elements, of which only n(n-1)/2 need to be calculated because of symmetry; likewise ν ist a totally symmetric tensor with at most n(n-1)(n-2) non vanishing elements, of which only n(n-1)(n-2)/6 need to be calculated. Therefore, all coefficients and thus u(x) can be calculated in p.t.

Take $F_5$ as an example. One gets

(11)    $8u(x) = 12 + x_1 - 2x_2 + 2x_3 - 3x_4 - x_1x_2 + x_1x_3 + x_2x_4 - x_3x_4 - x_1x_2x_3 - x_2x_3x_4$

For x=(1,1,1,1), e.g., (11) yields u(x)=1 in accordance with the formula F.

One can deduce some necessary conditions for the satisfiability of F from (10).

First: since u ≥ 0 and C > 0 the right hand side of u-C (see (10a))

$$\text{Rhs} := -\sum_{j=1}^{n} \lambda_i x_i + \sum_{<ij>} \mu_{ij} x_i x_j - \sum_{<ijk>} \nu_{ijk} x_i x_j x_k$$

cannot be smaller than –C and in order to make u vanish must compensate C exactly. Considered as a function of x in the hypercube $D = \{x \in \mathbb{R}, |x_i| \leq 1\}$ the largest possible value of Rhs is bounded by

$$\text{Rhsmax} = \sum_{j=1}^{n} |\lambda_i| + \sum_{<ij>} |\mu_{ij}| + \sum_{<ijk>} |\nu_{ijk}|$$

So if Rhsmax < C, F cannot be satisfied:

*Lemma VI*

If  $\sum_{j=1}^{n} |\lambda_i| + \sum_{<ij>} |\mu_{ij}| + \sum_{<ijk>} |\nu_{ijk}| < C$  then F is non-satisfiable.

To deduce another criterion I choose different weight factors $\alpha_i$, namely set all $c_i$=1. Then the constant term in u is just the number of clauses, m. All terms in u are whole numbers and u changes by an even amount if a configuration x is changed by reversing just one $x_i$. Thus – in order to compensate m exactly – the terms in Rhs must sum up to an even number if m is even, to an odd number if m is odd. Therefore

*Lemma VIIa*

If F is SAT  then u({$x_i$=1}) = $m - \sum_{j=1}^{n} \lambda_j + \sum_{<ij>} \mu_{ij} - \sum_{<ijk>} \nu_{ijk}$ = 0 mod 2

and vice versa:

*Lemma VIIb*

If u({$x_i$=1}) = 1 mod 2   then  F is non SAT

Note that all checks deduced so far are computable in p.t..

An additional check can be deduced by writing

$$u(x) = \sum_{i,j=1}^{n} M_{ij} x_i x_j - \sum_{j=1}^{n} \lambda_i x_i - \sum_{<ijk>} \nu_{ijk} x_i x_j x_k$$

with a (n,n)-matrix $M_{ij} = \frac{C}{n}\delta_{ij} + \frac{1}{2}\mu_{ij}$. Now calculate the lowest und largest eigenvalue of M, $e_{min}$ and $e_{max}$, respectively. Then by Raileighs criterion for quadratic forms the following bounds hold for the number of clauses unsatisfied by x:

$$ne_{min} - \sum_{j=1}^{n} \lambda_i x_i - \sum_{<ijk>} \nu_{ijk} x_i x_j x_k \leq u(x) \leq ne_{max} - \sum_{j=1}^{n} \lambda_i x_i - \sum_{<ijk>} \nu_{ijk} x_i x_j x_k,$$

and thus

> *Lemma VIIIa*:
>
> If $ne_{min} - \sum_{j=1}^{n} \lambda_i x_i - \sum_{<ijk>} \nu_{ijk} x_i x_j x_k > 0$ for all $x \in \{-1,1\}^n$, then F is non-satisfiable
>
> *Lemma VIIIb*:
>
> If $ne_{max} - \sum_{j=1}^{n} \lambda_i x_i - \sum_{<ijk>} \nu_{ijk} x_i x_j x_k < 1$ for some $x \in \{-1,1\}^n$, then F is satisfiable

To simplify the checks further we extend F to get rid of the cubic terms. Let F be a 3-SAT CNF-formula as before.

> *Definition*:
>
> $F_{ext}$ is called an *extension* of F iff $F_{ext}$ contains additional clauses such that *all* $\nu_{ijk}$ calculated for $F_{ext}$ *vanish*.

Such an extension can always be found by adding for each clause $(l_i, l_j, l_k)$ a clause with adverse "parity", e.g. $(l_i, \overline{l_j}, l_k)$ or any of the other three possibilities. Clearly, the construction can be done in polynomial time. This transformation does not conserve satisfiability, but the following holds:

> *Lemma IX*:
>
> If $F_{ext}$ is SAT then F is SAT

This is so because any solution of $F_{ext}$ necessarily fulfills all clauses of F, thus is a solution of F, too. Of course, the space of *all* solutions of F is not necessarily conserved.

Should $F_{ext}$ turn out UNSAT, however, nothing can be said about F, because the unsatisfiability of $F_{ext}$ may result from the additional clauses. The good thing is that satisfiability of $F_{ext}$ is easier to check, because u(x) for $F_{ext}$ is represented by a *quadratic form* in x.

Is there always an extension of F (for arbitrary F) which is satisfiable iff F is? The answer is yes. But in order to construct one at least one solution of F must be known. Alternatively one has to check out all possible $F_{ext}$. Therefore also this approach is not conclusive in determining whether 3-SAT is P or NP, because both alternatives may require exponential search depth.

**Satisfiability as a minimization problem.**

u(x) (or any corresponding quantity calculated with different $\alpha_i$ ) represents all features of F to determine its satisfiability. The task is twofold: determine the minimum of u(x) with respect to all possible assignments x, and see if the minimum value is zero. If the latter is the case solutions exist and F is satisfiable. Otherwise F is non-satisfiable.

Since u(x) is linear in each variable the following minimization procedure is at hand. Choose any of the $2^n$ variables $x_i$, say $x_1$. The remaining variables are combined in a vector y. Then write

$$(12) \quad u(x_1, y) = u(0, y) + x_1(-\lambda_1 + \sum_j \mu_{1j} x_j - \sum_{<jk>} \nu_{1jk} x_i x_j)$$

The factor

$$(13) \quad S_1(y) = -\lambda_1 + \sum_j \mu_{1j} x_j - \sum_{<jk>} \nu_{1jk} x_i x_j$$

of $x_1$ does not depend on $x_1$. From the definition of u(x), see equ. (5) and (9), it is clear that $u(0,y) \geq 0$. Thus the minimum of u is achieved only if $x_1 S_1 \leq 0$. Whether the minimum location has $x_1=1$ or $x_1=-1$ is determined by the sign of $S_1$:

$$(14) \quad u(1, y) - u(-1, y) = 2S_1(y)$$

We discriminate three cases:

(i)     $S_1(y) \geq 0$     (ii)     $S_1(y) \leq 0$     (iii)     $S_1(y)$ has positive and negative values

Note that the decision between (i) –(iii) can be done in p.t. for a READ-3 formula, because $S_1$ contains at most 6 different $x_j$ in that case. So $\min_y S_1(y)$ and $\max_y S_1(y)$ can be determined in a finite number of steps, independent of n or m.

In case (i) holds, according to (14) the location of the minimum of u has $x_1=-1$:

$$\min_x u(x) = \min_y u(-1, y)$$

whereas in case (ii) the opposite holds:

$$\min_x u(x) = \min_y u(1, y)$$

Note that u(1,y) and u(-1,y) are the functions defined by (5) and (9) belonging to a formula F' in which the variable $x_1$ has been assigned the value 1 or -1, respectively. If F was 3-SAT and READ-3 so is F'. So the problem of finding the minimum value of u is reduced by one variable in p.t. steps. Repeating the same procedure for the other variables finally leads to the minimum value of u and the decision whether F is satisfiable ($u_{min}$=0) or not.

Unfortunately there is condition (iii). Let $L_+ = \{y \mid S_1(y) \geq 0\}$ and $L_- = \{y \mid S_1(y) \leq 0\}$. Then

$$\min_x u(x) = \min\{\min_{y \in L_+} u(-1, y), \min_{y \in L_-} u(1, y)\}$$

Since it leads to branching the p.t. character of the procedure is disturbed.

To illustrate the fortunate way, consider the example $F_5$. From (11)

$$S_1(y) = (1 - x_2 + x_3 - x_2 x_3)/8$$

which can take values 0 and 1/2 for $x_2, x_3 \in \{-1, 1\}$. Thus the minimum must have $x_1$=-1, and one repeats the procedure for

$$8u(-1, y) = 11 - x_2 + x_3 - 3x_4 + x_2 x_4 - x_3 x_4 + x_2 x_3 - x_2 x_3 x_4$$

One may procede with

$$S_2(y) = (-1 + x_4 + x_3 - x_3 x_4)/8 \leq 0$$

to choose $x_2$=1, procede with u(-1,1,y) and so on to finally arrive at $u_{min}$=8.

There is a shortcut however, opened up by lemma VI in the preceding section: Since the x-terms in 8u(-1,y) cannot compensate the constant term 11, it is clear already that u cannot become zero and thus $F_5$ is unsatisfiable.

A simple example for which the minimization procedure outlined above does not run efficiently is

$$G = \begin{array}{ccccc} x & x & x & 0 & 0 \\ 0 & \bar{x} & \bar{x} & \bar{x} & 0 \\ 0 & 0 & x & x & \bar{x} \\ x & 0 & 0 & x & x \\ \bar{x} & x & 0 & 0 & x \end{array}$$

G is a 3-SAT, READ-3 formula and obviously not HORNSAT or trivial in the sense of lemmata Ia and Ib. G has 14 solutions as can be calculated easily from the formula for N(F), equ. (7). But all $S_i$, calculated for the 5 variables can take positive and negative values. In this case extension, as introduced in the preceding section, see lemma IX, can help. If we extend G to G' by adding the following clauses

| | | | | |
|---|---|---|---|---|
| $\bar{x}$ | x | x | 0 | 0 |
| 0 | x | $\bar{x}$ | $\bar{x}$ | 0 |
| 0 | 0 | $\bar{x}$ | x | $\bar{x}$ |
| $\bar{x}$ | 0 | 0 | x | x |
| x | x | 0 | 0 | x |

then the resulting G' has vanishing $\nu_{ijk}$'s and a considerably simpler u:

$$8u' = 10 - 4x_2 - 2x_4 - 2x_5 + 2x_2x_3 + 2x_2x_5 + 2x_3x_4$$

Now the minimization procedure works without branching ( $S_2 = -4 + 2x_3 + 2x_5 \leq 0$ ) and leads to a solution. Since a solution of G' fulfills all clauses, in particular those of the original G, G is determined to be satisfiable.

**Similarity with models of statistical physics.**

From (10) it is evident that u(x) can be considered as the hamiltonian of a dilute spin glass chain ($x_i = \pm 1$) with arbitrary longe range interactions ($\mu$) and a triple-spin interaction ($\nu$) in a locally varying magnetic field ($\lambda$). Thus the task of determining whether a given CNF-formula F is satisfiable is equivalent to the task of determining the – possibly highly degenerate – ground state of the corresponding "hamiltonian": if the ground state has zero energy F is satisfiable, otherwise F is non-satisfiable. The similarity of Boolean satisfiability with problems in statistical physics is the basis for intricate studies of the nature of the satisfiability via thermodynamic methods borrowed from the physics of spin glasses.

On the basis of this (not rigorously established analogy) one may ask the P=NP? question anew. As claimed by Barahona [8] both the twodimensional Ising-model with magnetic field and the threedimensional Ising model are NP-complete. Considered as a model of statistical physics the u(x)-representation of the satisfiability question is expected to have the same degree of complexity as a threedimensional Ising model because of the triple-spin interactions. This in conjunction with Barahonas result can be taken as an indication for SAT being NP.

On the other hand one can map (in polynomial time) a finite three-dimensional Ising-model to a 2-SAT problem of propositional calculus [9], which means that the question whether the spin system can occupy a state corresponding to the absolute minimum u=0 can be calculated in polynomial time, and thus the Ising model is of complexity P. Either Barahonas result is questionable or the analogy is too vague.